\documentclass[preprint,aps,nofootinbib,preprintnumbers,amsmath,amssymb,11pt]{revtex4}
\usepackage{epsfig}
\graphicspath{ {images/} }
\usepackage{ytableau}
\usepackage{mwe}
\usepackage{amsmath}
\usepackage{multirow}
\usepackage{slashed}
\usepackage{xcolor}
\usepackage{amssymb}
\usepackage{color}
\definecolor{darkblue}{rgb}{0.,0.,0.4}
\definecolor{darkred}{rgb}{0.5,0.,0.}
\definecolor{BlueViolet}{RGB}{138,43,226}
\definecolor{SkyBlue}{RGB}{30,144,255}
\definecolor{DarkGreen}{RGB}{0,100,0}
\usepackage[colorlinks=true,linkcolor=blue,citecolor=blue]{hyperref}
\usepackage{amsfonts}
\usepackage{graphicx}
\usepackage{dcolumn}
\usepackage{bm}
\usepackage{natbib}

\newcommand{\be}{\begin{equation}}
\newcommand{\ee}{\end{equation}}
\newcommand{\bea}{\begin{eqnarray}}
\newcommand{\eea}{\end{eqnarray}}

\newcommand{\nn}{\nonumber}

\newcommand{\imineq}[2]{\vcenter{\hbox{\includegraphics[height=#2ex]{#1}}}}

\def\im{{{\rm i}}}
\begin{document}

\preprint{DESY 21-114}

\title{Seeking SUSY fixed points in the $4-\epsilon$ expansion
}
\author{\vspace*{1 cm} Pedro Liendo}
\email{pedro.liendo@desy.de}
\author{Junchen Rong}
\email{junchen.rong@desy.de}
\affiliation{\vspace*{0.5 cm}DESY Hamburg, Theory Group, Notkestraße 85, D-22607 Hamburg, Germany\vspace*{2 cm}}

\begin{abstract}
\vspace*{1 cm}
We use the $4-\epsilon$ expansion to search for fixed points corresponding to $2+1$ dimensional $\mathcal{N}$=1 Wess-Zumino models of $N_{\Phi}$ scalar superfields interacting through a cubic superpotential. 
In the $N_{\Phi}=3$ case we classify all SUSY fixed points that are perturbatively unitary.
In the $N_{\Phi}=4$ and $N_{\Phi}=5$ cases, we focus on fixed points where the scalar superfields form a single irreducible representation of the symmetry group (irreducible fixed points). For $N_{\Phi}=4$  we show that the S5 invariant super Potts model is the only irreducible fixed point where the four scalar superfields are fully interacting.
For $N_{\Phi}=5$, we go through all Lie subgroups of O(5) and then use the GAP system for computational discrete algebra to study finite subgroups of O(5) up to order 800. This analysis gives us three fully interacting irreducible fixed points. 
Of particular interest is a subgroup of O(5) that exhibits O(3)/Z2 symmetry. 
It turns out this fixed point can be generalized to a new family of models, with $N_{\Phi}=\frac{\rm N(N-1)}{2}-1$ and O(N)/Z2 symmetry, that exists for arbitrary integer N$\geq3$. 
\end{abstract}

\vspace*{3 cm}

\maketitle

\def\thesection{\arabic{section}}
\def\thesubsection{\arabic{section}.\arabic{subsection}}
\numberwithin{equation}{section}
\newpage

\tableofcontents
\section{Introduction}
In recent years, $2+1$ dimensional superconformal field theories (SCFTs) with minimal ($\mathcal{N}=1$) supersymmetry have received significant attention \cite{Grover:2013rc,Bashkirov:2013vya,Fei:2016sgs,Bashmakov:2018wts,Benini:2018bhk,Gaiotto:2018yjh,Benini:2018umh,Rong:2018okz,Atanasov:2018kqw,Rong:2019qer}.
Perhaps the simplest model with this symmetry is the so-called $\mathcal{N}=1$ super-Ising model, which was proposed in \cite{Grover:2013rc}  to describe a quantum critical point at the boundary of a $3+1$ dimensional topological superconductor. Because the super-Ising model preserves time reversal symmetry, its spectrum contains only one relevant operator which is time-reversal invariant. This means that a generic renormalization group flow (not necessarily supersymmetric) can reach the fixed point by tuning a single parameter. This property is called emergent supersymmetry and might be realizable in experiment. 


Apart from the super-Ising, a whole zoo of theories have been identified and they are sometimes related by  dualities \cite{Bashmakov:2018wts, Benini:2018bhk,Gaiotto:2018yjh,Benini:2018umh}. In particular, a certain super QED was shown to be dual to an $\mathcal{N}=1$ Wess-Zumino model \cite{Benini:2018bhk,Gaiotto:2018yjh}. This duality has many proprieties that resemble the self duality of non-supersymmetric
QED coupled to two complex scalars, which describes the famous de-confinement quantum critical point \cite{senthil2004deconfined,wang2017deconfined}.   

The study of $2+1$ dimensional $\mathcal{N}=1$ SCFT has also been boosted by technical improvements on standard techniques.
For example, it was observed in \cite{Fei:2016sgs} that if one sets the number of Dirac fermions to $N_f=\frac{1}{4}$ in four dimensions, after analytic continuation to $4-\epsilon$ dimensions, one ends up with a single three-dimensional Majorana fermion when $\epsilon=1$. 
By studying the Gross-Neveu-Yukawa model with $N_f=\frac{1}{4}$ Dirac fermions and one real scalar, one gets a perturbative fixed-point which is in good agreement with the $\mathcal{N}=1$ super-Ising model in $2+1$ dimensions. 
This technique can be easily generalized to more general $2+1$ dimensional $\mathcal{N}=1$ Wess-Zumino models with superpotential\footnote{The quadratic terms break time-reversal symmetry and cannot be added to the superpotential (see for example \cite{Gaiotto:2018yjh}).}
\be\label{N1superpotential}
\mathcal{W}=\frac{1}{6}h_{ijk}\Phi^i\Phi^j\Phi^k\,.
\ee
In other words, one can study the corresponding Gross-Neveu-Yukawa model
\be
\mathcal{L}_{\mathcal{N}=1}=\frac{1}{2}(\partial\phi^i)^2+\frac{1}{2}\bar{\psi^i}\partial\cdot \Gamma \psi^i+\frac{1}{2} h_{ijk} \phi^i\bar{\psi^j}\psi^k+ \frac{1}{8} h_{ijm}h_{klm}\phi^i\phi^j\phi^k\phi^l\,,
\ee
with $N_f=\frac{N_{\Phi}}{4}$.

Parallel to these analytical developments, progress has also been made using the highly successful numerical bootstrap program \cite{Rattazzi:2008pe,Poland:2018epd}. 
Recent works that studied  $2+1$ dimensional $\mathcal{N}=1$ SCFTs using numerical techniques include \cite{Bashkirov:2013vya,Rong:2018okz,Atanasov:2018kqw,Rong:2019qer}. 
In particular, when applied to the $\mathcal{N}=1$ super-Ising model, the scaling dimension of the superfield $\Phi$ can be determined to very high precision \cite{Rong:2018okz}. 

In this work we would like to find an organizational scheme for three-dimensional models with $\mathcal{N}=1$ supersymmetry. 
Our analysis will be perturbative and rely in the standard $\epsilon$-expansion. 
The logic being that a one-loop analysis should give us at least a qualitative understanding of potential fixed points. Once we know a fixed point exists, and have a basic idea of its spectrum, we can use more powerful techniques like the modern numerical bootstrap in order to solve it to high precision. This work is then similar to \cite{Osborn:2017ucf,Codello:2019isr,Osborn:2020cnf}, where scalar models in several dimensions were studied using a similar method.

Our approach is then the following, we use the $4-\epsilon$ expansion to search for $\mathcal{N}=1$ SCFTs with a small number of scalar superfields interacting through the superpotential \eqref{N1superpotential}. 
We systematically study all the cases when the number scalar superfields $N_{\Phi}$ is less or equal than 5. 
In Section \ref{threecomponent}, we start with the $N_{\Phi}=3$ case, which will also include all the possible solutions for $N_{\Phi}=1,2$. For  $N_{\Phi}=3$ the superpotential has 10 independent couplings $h_{ijk}$, by solving the one-loop beta functions we list in Table \ref{N3points} all the possible fixed points with $N_{\Phi} \leq 3$  that are perturbatively unitary. 
We then increase the number of scalar to $N_{\Phi}=4$ in Section \ref{fourcomponent} and to $N_{\Phi}=5$ in Section \ref{fivecomponents}. We focus on the SUSY fixed points where the scalar superfields form a single irreducible representation of the symmetry group. We call these fixed points ``irreducible fixed points''. 
This means the coupling constants $h_{ijk}= \sum_a g_a d^{a}_{ijk}$ are a linear superposition of the degree three invariant tensors of the symmetry group. 
The invariant tensors are traceless and also satisfy
\be
d^{a}_{imn}d^{b}_{jmn}\propto \delta_{ij}\,.
\ee
This is essentially the trace condition used in \cite{Brezin:1973jt}, and has important physical consequences. 
It implies there is only one boson mass operator $\phi_i\phi_i$ that is invariant under the symmetry group of the fixed points. 
For the most stable fixed point of the beta function, this operator is also the only relevant operator that preserves both time-reversal symmetry and the flavor symmetry of the SCFT. 
Similar to the super-Ising model, non-supersymmetric RG flows will reach the supersymmetric fixed point as long as one tunes the coupling of the boson mass term to zero. In other words, these systems have emergent supersymmetry. 
The results of the classification are listed in Table \ref{N4N5points}. 
In the $N_{\Phi}=4$ case, it turns out that there is only one group that gives us a fixed point where the four superfields are fully interacting. 
This result is based on the analysis of all the subgroups of O(4) in \cite{michel1981landau}. 
In the $N_{\Phi}=5$ case, we focus on Lie subgroups and finite subgroups of O(5) with orders less or equal to 800.  
In Section \ref{generalN}, we observe that some fixed points for low values of  $N_{\Phi}$ can be considered the first entry in a family of fixed points that can be generalized to higher values of $N_{\Phi}$. For example, the O(3)/Z2 SCFT presented in Table \ref{N4N5points} can be generalized to a new family of SCFTs preserving the O(N)/Z2 symmetry. This family complements the three families (the super Potts models, the SU(N) invariant SCFTs and the F4-family of SCFTs) of Wess-Zumino models already studied in \cite{Rong:2019qer}.

\begin{table}[ht]
\begin{tabular}{|c|c|c|c|c|}
\hline
$N_{\Phi}$                & $v$ and $\lambda$'s & $\gamma_{\Phi}$                                                                                                                  & Symmetry         & $A/\epsilon^2$ \\ \hline\hline
1                  & \eqref{sol1}        & $\frac{1}{14} \epsilon$                                                                                                          &                  & -19.7392       \\ \hline\hline
\multirow{2}{*}{2} & \eqref{sol2}        & $\frac{1}{6} \epsilon, \frac{1}{6} \epsilon$                                                                                     & S3               & -92.1163       \\ \cline{2-5} 
                   & \eqref{sol3}        & $\frac{9}{109} \epsilon$, $\frac{13}{218}\epsilon$                                                                               & Z2               & -39.2973       \\ \hline\hline
\multirow{6}{*}{3} & \eqref{sol4}        & $\frac{1}{10} \epsilon$, $\frac{1}{10} \epsilon$, $\frac{1}{10} \epsilon$                                                        & S4               & -82.9047       \\ \cline{2-5} 
                   & \eqref{sol5}        & $0.232069 \epsilon$, $0.146209\epsilon$, $0.146209\epsilon$                                                                      & O(2)             & -144.941       \\ \cline{2-5} 
                   & \eqref{sol6}        & $\frac{9}{190}\epsilon,  \frac{47}{570}\epsilon,  \frac{47}{570}\epsilon$                                                        & S3               & -58.6635       \\ \cline{2-5} 
                   & \eqref{sol7}        & $\frac{49}{290}\epsilon$, $\frac{1}{290} \left(\sqrt{265}+29\right)\epsilon$, $\frac{1}{290} \left(29-\sqrt{265}\right)\epsilon$ & K4=Z2$\times$ Z2 & -110.647       \\ \cline{2-5} 
                   & \eqref{sol8}        & $0.171857\epsilon, 0.16459\epsilon, 0.063942\epsilon $                                                                           & Z2               & -58.8549       \\ \cline{2-5} 
                   & \eqref{sol9}        & $0.0871871\epsilon, 0.0710876\epsilon, 0.0546982\epsilon$                                                                        & Z2               & -101.963       \\ \hline
\end{tabular}
\caption{All perturbatively unitary $\mathcal{N}$=1 fixed points with three or less interacting scalar superfields.}
\label{N3points}
\end{table}

\begin{table}[h]
\begin{tabular}{|c|c|c|c|l|}
\hline
$N_{\Phi}$                 & $\gamma_{\Phi}$        & Symmetry & $A/\epsilon^2$ &                                         \\ \hline\hline
4                  & $\frac{3}{34}\epsilon$ & S5       & -97.5349       & Section \ref{S5potts}                   \\ \hline\hline
\multirow{3}{*}{5} & $\frac{1}{12}\epsilon$ & S6       & -115.145       & \multirow{3}{*}{Section \ref{N5points}} \\ \cline{2-4}
                   & $\frac{3}{20}\epsilon$ & S5       & -207.262       &                                         \\ \cline{2-4}
                   & $\frac{7}{30}\epsilon$ & O(3)/Z2       & -322.407       &                                         \\ \hline 
\end{tabular}
\caption{Irreducible fixed points of $N_{\Phi}$=4 and $N_{\Phi}$=5. The N=4 fixed point listed here is the only irreducible fixed point of four fully interacting real scalars superfields. The $N_{\Phi}$=5 fixed points listed here are the only irreducible fixed points of five fully interacting real scalars superfields that have a finite symmetry whose order is less than 800. }\label{N4N5points}
\end{table}

\section{General proprieties of the $\mathcal{N}=1$ SCFTs}
The one-loop beta function of the $\mathcal{N}=1$ SCFT Wess-Zumino model with $N_{\Phi}$ scalar superfields coupled through the superpotential \eqref{N1superpotential} is given by \cite{Fei:2016sgs},
\be
\beta(h)=-\frac{\epsilon}{2} h_{ijk}+\frac{1}{16\pi^2}(\frac{1}{2}(h_{ijm}h_{kpq}h_{mpq}+h_{ikm}h_{jpq}h_{mpq}+h_{jkm}h_{ipq}h_{mpq})+2 h_{imp}h_{jpq}h_{kqm})\, ,
\ee
and the anomalous dimension matrix (whose eigenvalues give us the anomalous dimension) reads
\be
(\gamma_{\Phi})_{ij}=\frac{1}{16\pi^2}\frac{1}{2}h_{imn}h_{jmn}\, .
\ee
Like the scalar theory in $6-\epsilon$ dimensions, this is a gradient flow \cite{Grinstein:2014xba,Osborn:2017ucf,Codello:2019isr}. In other words, 
\be
\beta_{ijk}(h)=\partial A/\partial h_{ijk}
\ee
with 
\be
A=-\epsilon h_{ijk}h_{ijk}+\frac{1}{16\pi^2}(\frac{3}{8}h_{ijm}h_{kpq}h_{mpq}h_{ijk}+\frac{1}{2}h_{imp}h_{jpq}h_{kqm}h_{ijk})\,.
\ee
Along the RG flow to the IR, $A$ decreases and at the fixed points, 
\be
A=-\frac{7\epsilon}{8}h_{ijk}h_{ijk}=-28 \pi ^2 \epsilon \sum_{i}\gamma_i\,.
\ee
Here $\gamma_i$ are the eigen-values of the anomalous dimension matrix.

\section{$N_{\Phi}=3$ reducible fixed points}\label{threecomponent}
The strategy we employ in this section is identical to the one described in \cite{Codello:2019isr} to study scalar $\phi^3$ theory in $6-\epsilon$ dimensions. 
Notice that the coupling constant $h_{ijk}$ is a fully symmetric tensor and can be characterized as
\be
h_{ijk}=v_{(i}\delta_{jk)}+\lambda_{ijk}\,.
\ee
Here $\lambda_{ijk}$ is symmetry and traceless. 
We parameterize the potential as 
\bea
\mathcal{W}(\Phi)&=&(v_j\Phi_j)(\Phi_i\Phi_i)-\frac{\lambda _1 \Phi _1 \left(\Phi _1^2-3 \Phi _2^2\right)}{2 \sqrt{2}}+\frac{\lambda _2 \Phi _2 \left(\Phi _2^2-3 \Phi _1^2\right)}{2 \sqrt{2}}
+\frac{1}{2} \sqrt{3} \lambda _3 \left(\Phi _2^2-\Phi _1^2\right) \Phi _3\nonumber\\&&+\sqrt{3} \lambda _4 \Phi _1 \Phi _2 \Phi _3-\frac{1}{2} \sqrt{\frac{3}{10}} \lambda _5 \Phi _1 \left(\Phi _1^2+\Phi _2^2-4 \Phi _3^2\right)-\frac{1}{2} \sqrt{\frac{3}{10}} \lambda _6 \Phi _2 \left(\Phi _1^2+\Phi _2^2-4 \Phi _3^2\right)\nonumber\\&&+\frac{\lambda _7 \Phi _3 \left(-3 \Phi _1^2-3 \Phi _2^2+2 \Phi _3^2\right)}{\sqrt{10}}\,.
\eea
The parameterization relies on the representation theory of O(3). 
The couplings $$\lambda_1+\im \lambda_2, \lambda_3+\im \lambda_4, \lambda_5+\im \lambda_6, \lambda_7, \lambda_5-\im \lambda_6, \lambda_3-\im \lambda_4 \text{ and } \lambda_1-\im \lambda_2$$ form a $j=3$ representation of O(3) with $m=+3,+2,+1,0,-1,-2,-3$ respectively. Clearly, $v_j$ forms a vector representation of O(3). We can use the $J_1$ and $J_2$ rotation to set $v_1=v_2=0$,and after that we can use $J_3$ to set $\lambda_1=0$. 
After fixing the redundant O(3) rotations solution of $\beta(v,\lambda)=0$ are discrete. The super Ising model plus two decoupled free scalar superfields is located at 
\be\label{sol1}
v_3=-\frac{2 \pi }{5 \sqrt{7}},\quad  \lambda_7=\frac{-2}{3}  \sqrt{\frac{2}{35}} \pi\,, 
\ee
and all other $v$'s and$ \lambda$'s vanishing. 

The S3 invariant fixed point (S3 super Potts model) of two scalars plus a decoupled scalar superfield is located at 
\be\label{sol2}
\lambda_1=\frac{4 \pi }{3 \sqrt{3}},
\ee
and all other $v$'s and$ \lambda$'s vanishing. The Lagrangian of the $S_3$ super Potts model can in fact be re-written as an $\mathcal{N}$=2 supersymmetric Lagrangian with a single complex chiral superfield \cite{Rong:2018okz}. This explains the anomalous dimension $\gamma_{\Phi}=\frac{1}{6}\epsilon.$

The Z2 fixed point with two coupled scalars plus one decoupled scalar superfield is located at 
\be\label{sol3}
v_3= \frac{2 \pi }{\sqrt{109}},\quad \lambda_4=2 \sqrt{\frac{3}{109}} \pi,\quad \lambda_7=-\frac{1}{3} \sqrt{\frac{10}{109}} \pi,
\ee
and all other $v$'s and$ \lambda$'s vanishing. 

The S4 invariant super Potts model is located at
\be\label{sol4}
\lambda_3=2 \sqrt{\frac{2}{15}} \pi,
\ee
and all other $v$'s and$ \lambda$'s vanishing. 

The O(2) fixed point with the three scalars fully coupled is located at
\be\label{sol5}
v_3=0.440267,\quad \lambda_7=-2.06839,
\ee
and all other $v$'s and$ \lambda$'s vanishing. 

The S3 invariant fixed point with all three scalars fully coupled is located at
\be\label{sol6}
v_3=\frac{2 \pi }{\sqrt{95}},\quad \lambda_1=\frac{4}{3} \sqrt{\frac{23}{285}} \pi, \quad \lambda_7 =\frac{-2}{3}  \sqrt{\frac{2}{19}} \pi,
\ee
and all other $v$'s and$ \lambda$'s vanishing. 

The K4 invariant fixed point with all three scalars fully coupled is located at
\be\label{sol7}
v_3=-\frac{2 \pi }{5 \sqrt{29}},\quad \lambda_4=-2 \sqrt{\frac{53}{435}},\quad \lambda_7=\frac{-7}{3}  \sqrt{\frac{2}{145}} \pi
\ee
and all other $v$'s and$ \lambda$'s vanishing. 

The Z2 invariant fixed point with all three scalars fully coupled is located at
\be\label{sol8}
v_3=0.417419,\quad \lambda_1=-2.19206,
\quad \lambda_3=0.962964,\quad \lambda_5=0.401921,\quad\lambda_7=0.524619.
\ee
and all other $v$'s and$ \lambda$'s vanishing. .

Finally, another Z2 invariant fixed point with all three scalars fully coupled is located at
\bea\label{sol9}
&& v_3=0.708567,\quad \lambda_1=-1.08699, \quad \lambda_3=0.0194906,\nonumber\\&& \lambda_4=0.0337588,\quad \lambda_5=0.13039,\quad \lambda_6=0.225843,\quad \lambda_7=-0.63725.
\eea

A summary of the previous analysis is presented in Table~\ref{N3points}.
We can compare our results with the perturbative fixed points of scalar $\phi^3$ theory in $6-\epsilon$ dimension, which were studied in \cite{Codello:2019isr}. Our entries of Table~\ref{N3points} turn out to be in one-to-one correspondence with the fixed points of scalar $\phi^3$ theory in $6-\epsilon$ dimensions. In $6-\epsilon$ dimensions, only the O(2) fixed point and the K4 fixed point are perturbatively unitary. In our case, it seems that supersymmetry improves the situation a lot, and all the fixed points become perturbatively unitary.   

\section{$N_{\Phi}=4$ irreducible fixed points}
\label{fourcomponent}

We now discuss the case when the scalar superfield transforms as an irreducible representation of the symmetry group. To construct the superpotential \eqref{N1superpotential}, the symmetry group needs to preserve a rank-3 fully symmetric invariant tensor $d^a_{ijk}$. 
The tensors $d^a_{ijk}$ are necessarily traceless, such that the representation is reducible. Another important condition is that the invariant tensors satisfy\footnote{Otherwise one can build projectors of the form
\be
 d_{imn}d_{jmn}-e_1 \delta_{ij},
\ee which projects the vector representation into invariant subspace. Here $e_1$ is the first eigenvalue of $d_{imn}d_{jmn}$.}
\be
d^{a}_{imn}d^{b}_{jmn}\propto \delta_{ij}\,.
\ee
This is analogous to the trace condition used in \cite{Brezin:1973jt}.
We can now normalize $d^{a}_{ijk}$ to satisfy 
\be\label{nomal}
d^{a}_{imn}d^{b}_{jmn}=\delta_{ab} \delta_{ij}\,.
\ee

If the symmetry group preserves a single $d_{ijk}$, the beta function and the anomalous dimension become 
\bea\label{betaandgamma}
&&\beta_{g}=-\frac{g}{2}\epsilon+\frac{4 T_3+3}{32 \pi ^2}g^3,\nonumber\\
&&\gamma_{\Phi}=\frac{\epsilon }{8 T_3+6}.
\eea
Here $T_3$ is defined through 
\be\label{T3}
d_{imn}d_{jmp}d_{kpm}=T_3  d_{ijk},
\ee
assuming $d_{ijk}$ satisfies the normalization \eqref{nomal}.
Using the fact that 
\be
\sum_{ijkl}(d_{ijm}d_{klm}\pm d_{ikm}d_{jlm})^2\geq0\,,
\ee
one can prove that
\be
-1\leq T_3\leq 1\,.
\ee
From the beta function \eqref{betaandgamma}, we know that the one-loop fixed point exists if and only if
\be\label{condtion}
T_3>-\frac{3}{4}\,.
\ee

The irreducible subgroups of O(4) were studied extensively in \cite{michel1981landau}.
It was discovered that only five of the groups have a four dimensional irreducible real representation that preserves a degree three polynomial \footnote{The representation is irreducible when the fields are real,  but can be reducible when they are complex. In that case, the representation equals the sum of two complex conjugate representations.}. 
The result is recalled in Table \ref{subgroup}. The five polynomials are 
\bea\label{poly4}
&&I_1=x(x^2-3y^2)+z(z^2-3 t^2)\,,\nonumber\\
&&I_2=y(y^2-3x^2)+t(t^2-3 z^2)\,,\nonumber\\
&&I_3=t^3-t(x^2+y^2+z^2)+10/\sqrt{5}x y z\,,\nonumber\\
&&I_4=x(z^2-t^2)+t(x^2-y^2)-2yz(x+t)\,,\nonumber\\
&&I_5=x (x^2-3y^2)-t(t^2-3z^2)\,.
\eea
\begin{table}[ht]
\begin{tabular}{|c|c|c|c|}
\hline SmallGroup Id & subgroup   & order               & invariant polynomial  \\\hline\hline
[120,34] & S5  & 120 & $I_3$ \\ \hline
[72,40] &  (S3$\times$S3)$\rtimes$ Z2  &72 & $I_1$\\ \hline
[36,10] & S3$\times$ S3  &36 & $I_5$\\ \hline
[20,3]& Z5$\rtimes$ Z4   &20& $I_4$\\ \hline
[18,4] & (Z3$\times$Z3)$\rtimes$ Z2  &18& $I_1$ and $I_2$ \\ \hline
\end{tabular}
\caption{Irreducible subgroup of O(4) and the invariant polynomial preserved. }\label{subgroup}
\end{table}
The corresponding fully symmetric invariant tensor can be easily constructed from these polynomials 
\be\label{invtensor}
d^{a}_{ijk}=\frac{\partial^3 I_a}{\partial x_i\partial x_j\partial x_k}\,, \quad \text{with}\quad  \vec{x}=\{x,y,z,t\}\,.
\ee{}

Of these five polynomials only two of them are independent. We can pick these two independent polynomials to be $I_3$ and $I_1$. The polynomial $I_4$ is related to $I_3$ by an O(3) rotation. The group that preserves $I_4$ is Z5$\rtimes$Z4, which is, in fact, a subgroup of S5 (which preserves $I_3$). The two groups have the same degree three polynomial but different degree four polynomials. This means that to break the symmetry from S5 to Z5$\rtimes$Z4, it is necessary to include quartic terms in the superpotential \eqref{N1superpotential}. 
Since we include only cubic terms (the quartic terms are not only irrelevant, but also break time-reversal symmetry), the symmetry of our Lagrangian will be S5. Also, $I_5$ is related to $I_1$ by an O(3) rotation. 
Including either $I_1$ or $I_5$ in the superpotential \eqref{N1superpotential} will preserve the symmetry (S3$\times$S3)$\rtimes$Z2. The subgroup (Z3$\times$Z3)$\rtimes$ Z2 preserves two polynomials $I_1$ and $I_2$. However,  any combination of the two polynomial $a I_1+b I_2$ can be brought back to the form $c I_1$ by an O(3) rotation (the rotation parameter depends on a and b). This means a superpotential of the form $\mathcal{W}=g_1 I_1+g_2 I_2$ in fact preserves the symmetry (S3$\times$S3)$\rtimes$Z2 at any of point of the $(g_1, g_2)$-plane. This of course is also true for the SCFT fixed point.

\subsection{The fixed points and anomalous dimensions}\label{S5potts}
We can now use the explicit form of the polynomial $I_3$ to study the S5 invariant fixed point. Plug $I_3$ into \eqref{invtensor}, re-scale it to satisfy the normalization \eqref{nomal}, and then plug it in \eqref{T3}, we get
\be
T_3=\frac{2}{3}\,.
\ee
Since $T_3>-\frac{3}{4}$, the fixed point exists, and from \eqref{betaandgamma} we know
\be
\gamma_{\Phi}=\frac{3}{34}\epsilon\,.
\ee
This model belongs to the family of $\cal{N}$=1 Potts models studied in \cite{Rong:2019qer}.

Similarly, we can also use the polynomial $I_1$ to study the (S3$\times$S3)$\rtimes$Z2 invariant fixed point, we get
\be
T_3=0\,.
\ee
Since $T_3>-\frac{3}{4}$, the fixed point exists, and we get
\be
\gamma_{\Phi}=\frac{1}{6}\epsilon\,.
\ee
Consider a superpotential $\mathcal{W}=g I_1$, we know that
\bea
\mathcal{W}&=& g (\phi^3+\chi ^3+ c.c.),
\eea
with 
\be 
\phi=x+\im y ,\quad \chi= z+\im t.
\ee
So that the model is simply two decoupled copies of the S3 super Potts model.

\section{
$N_{\Phi}=5$  irreducible fixed points}\label{fivecomponents}
\subsection{The Lie group O(3)}\label{O3Z2}
Among the Lie subgroups of O(5), there is only one subgroup where the 5 dimensional vector irrep remains irreducible.
This is the group O(3), under this embedding, the 5 of O(5) become the T irrep of O(3).\footnote{We denote the symmetric traceless, antisymmetric and singlet irrep of O(N) by T, A and S respectively.}
The invariant tensor of such a representation can be constructed using the $\delta_{ij}$ of O(3), we will postpone the details of the construction to Section \ref{ONTfixedpoints}, where we discuss the generalization of this fixed point to a series SCFTs with the scalar superfields transforming in the T irrep of O(N).
We note down here the $T_3$ constant,
\be
T_3=-\frac{3}{14}.
\ee
Since $T_3>-\frac{3}{4}$, the one-loop fixed point exists, the anomalous dimension is 
\be
\gamma_{\phi}/\epsilon=\frac{7}{30}\,,
\ee
and the corresponding A function reads
\be
A/\epsilon^2=-\frac{1}{3} \left(98 \pi ^2\right)\,.
\ee
Notice the Z2 transformation 
\be
\textrm{Z}=\left(
\begin{array}{ccc}
 -1 & 0 & 0 \\
 0 & -1 & 0 \\
 0 & 0 & -1 \\
\end{array}
\right),
\ee
acts trivially on the T irrep of O(3), so that the symmetry of the fixed point is 
\be
\textrm{O(3)/Z2}\,.
\ee

\subsection{Finite subgroups of O(5) up to order 800 with 5 dimensional faithful irreps}
\label{GAPsubgroups}

We now use the GAP system \cite{GAP4} for computational discrete algebra and the Small Groups library \cite{eick2018smallgrp} to search for finite groups with 5 dimensional faithful irreducible representations, similar to the study done in \cite{Ludl:2010bj}. Since all representations of finite groups are unitary representations, these finite groups are subgroups of U(5). 
There are two theorems that are especially useful in seeking faithful irreducible representations \cite{Ludl:2010bj}.
\begin{itemize}
    \item \textbf{Theorem 1} Suppose a finite group G has a p dimensional irreducible representation, then Ord(G)/p is an integer. 
    \item \textbf{Theorem 2} Suppose a finite group G has a p dimensional faithful irreducible representation, then this irreducible representation has a single character p in the character table.
\end{itemize}
The proof of \textbf{Theorem 1} can be found in \cite{hall2018theory} Page 288. The proof of \textbf{Theorem 2} was given in \cite{Ludl:2010bj} Appendix A.1. The SmallGroup library allows one to specify a finite group using two integers $[p,q]$,  the first integer indicates the order of the group,  while the second integer enumerates all finite groups with order $p$. 
We first select finite groups that are not Abelian. For example, the following command 
\be
\text{l := Filtered( AllSmallGroups(60) ,x - $>$ IsAbelian(x)=false);;}
\ee
gives all non-Abelian groups with order 60. The GAP system also allows us to calculate the character table easily
\be
\text{Display(CharacterTable(SmallGroup(60,5)));}.
\ee
To select 5 dimensional irreps, we can use
\be\label{cmd1}
\text{psi := Filtered( Irr( CharacterTable( SmallGroup(60,5)  ) ), x -$>$ Degree( x ) = 5);;}.
\ee
The symbol psi is now a list which contains all the characters of the 5 dimensional irreps:
\bea
&&\text{gap$>$ psi[1];}\nonumber\\
&&\text{$>$}\quad\quad\text{Character( CharacterTable( Alt( [ 1 .. 5 ] ) ), [ 5, -1, 1, 0, 0 ] )}
\eea
The characters contain a single 5, so according to Theorem 2 this is a faithful irrep. It is sometimes useful to check the structure description of the finite group \be
\text{StructureDescription(SmallGroup(60,5));}.
\ee
From the output, we learn that SmallGroup(60,5) is the alternating group of five elements, or ``A5''. Since our goal is to search for degree 3 invariant polynomials, we can also use the command 
\be
\text{MolienSeries(psi[1]); }
\ee
to obtain the Molien Series of the corresponding irrep. The Molien series
\be
M(z)=\frac{1}{|G|}\sum_{g\in G}\frac{1}{det[\mathbb{I}-z \rho(g)]}
\ee
is a generating function which counts the number of invariant polynomials of a certain degree. Here $|G|$ is the order of the group G, $\rho(g)$ is a representation of G. Doing a series expansion of $M(z)$, the coefficient of $z^n$ counts the number of invariant polynomials of degree n. The Molien series of the five-dimensional irrep of A5 is 
\be
\frac{z^{10}-z^8+z^6+z^5+z^4-z^2+1}{\left(z^2-1\right)^2 \left(z^3-1\right)^2 \left(1-z^5\right)}=1+z^2+2 z^3+2 z^4+O\left(z^5\right)\,.
\ee
So that it has one degree two invariant polynomial, two degree three polynomials and two degree four polynomials. 

There are more than $10^7$ finite groups of order less or equal than 800. Among them, we found only 109 groups that have at least one 5 dimensional faithful irrep.
They are listed in Table \ref{groups}. 
Among these 109 finite groups, it turns out 13 of them are also subgroups of O(5). That is, the finite groups preserve a bilinear form, or in other words, their Molien series takes the form
\be
1+z^2+O\left(z^3\right)\,.
\ee
These groups are marked with an ``*'' in Table \ref{groups}.
Among the 13 finite subgroups of O(5), only four of them have irreducible representations that preserve at least one degree three invariant polynomial.  
The results are listed in Table \ref{group5inv}. For a fixed $N_{\Phi}$, bigger groups tend to preserve less invariant polynomials, which leads us to conjecture that this list is complete.\footnote{We actually extended the analysis up to order 1000 and our final result does not change.}
\begin{table}[]
\begin{tabular}{|l|l|l|l|l|}
\hline
ID                             & Group               & Irrep     & Molien Series                                                                                                                                                          & Molien Series (expansion)                                                      \\ \hline\hline
{[}60, 5{]}                    & A5                  & 5         & $\frac{z^{10}-z^8+z^6+z^5+z^4-z^2+1}{\left(1-z^2\right)^2 \left(1-z^3\right)^2 \left(1-z^5\right)}$                                                                    & $1+z^2+2 z^3+2 z^4+O\left(z^5\right)$                                          \\ \hline
\multirow{2}{*}{{[}120, 34{]}} & \multirow{2}{*}{S5} & $5_a$     & $\frac{z^{12}+z^{10}+z^9+z^8+z^6+1}{\left(1-z^2\right) \left(1-z^3\right) \left(1-z^4\right) \left(1-z^5\right) \left(1-z^6\right)}$                                   & $1+z^2+z^3+2 z^4+O\left(z^5\right)$                                            \\ \cline{3-5} 
                               &                     & $5_b$     & $\frac{z^{14}-z^{13}+z^{12}-z^{11}+z^{10}-z^9+2 z^8-z^7+2 z^6-z^5+z^4-z^3+z^2-z+1}{(1-z) \left(1-z^3\right) \left(1-z^4\right) \left(1-z^5\right) \left(1-z^6\right)}$ & $1+z^2+z^3+2 z^4+O\left(z^5\right)$                                            \\ \hline
\multirow{2}{*}{{[}360,188{]}} & \multirow{2}{*}{A6} & $5_{std}$ & $\frac{z^{12}-z^9+z^6-z^3+1}{\left(1-z^2\right) \left(1-z^3\right)^2 \left(1-z^4\right) \left(1-z^5\right)}$                                                           &  $1+z^2+z^3+2 z^4+O\left(z^5\right)$ \\ \cline{3-5} 
                               &                     & 5'        & $\frac{z^{12}-z^9+z^6-z^3+1}{\left(1-z^2\right) \left(1-z^3\right)^2 \left(1-z^4\right) \left(1-z^5\right)}$                                                           & $1+z^2+z^3+2 z^4+O\left(z^5\right)$                                            \\ \hline
\multirow{2}{*}{{[}720,763{]}} & \multirow{2}{*}{S6} & $5_{std}$ & $\frac{1}{\left(1-z^2\right) \left(1-z^3\right) \left(1-z^4\right) \left(1-z^5\right) \left(1-z^6\right)}$                                                             & $1+z^2+z^3+2 z^4+O\left(z^5\right)$                                            \\ \cline{3-5} 
                               &                     & 5'        & $\frac{1}{\left(1-z^2\right) \left(1-z^3\right) \left(1-z^4\right) \left(1-z^5\right) \left(1-z^6\right)}$                                                             & $1+z^2+z^3+2 z^4+O\left(z^5\right)$                                            \\ \hline
\end{tabular}
\caption{Finite groups and their irreducible representation that preserves a degree two invariant polynomial and at least one degree three invariant polynomials. }
\label{group5inv}
\end{table}

\subsubsection{The group A5}
The SmallGroup([60,5]) is the alternating group of five elements.  As is explicit from the Molien series, it has a five dimensional irrep, which has a degree two invariant polynomial and two degree five polynomial. To work out the explicit form of the polynomial, after running command \eqref{cmd1}, we can use
\be
\text{IrreducibleRepresentationsDixon( SmallGroup(60,5),psi[1] );}
\ee
to calculate a matrix representation of the generators of the group
\be
\tilde{g}_1=(1,2,3,4,5)=\left(
\begin{array}{ccccc}
 0 & 1 & 0 & 0 & 0 \\
 -1 & 0 & -1 & 0 & -1 \\
 1 & -1 & 0 & -1 & 0 \\
 1 & 0 & 0 & 0 & 0 \\
 0 & 0 & 1 & 0 & 0 \\
\end{array}
\right), \quad \tilde{g}_2=(1,2,3,4,5)=\left(
\begin{array}{ccccc}
 0 & 0 & 1 & 0 & 0 \\
 0 & 0 & 0 & 1 & 0 \\
 -1 & 0 & -1 & 0 & -1 \\
 0 & -1 & 0 & -1 & -1 \\
 0 & 0 & 0 & 0 & 1 \\
\end{array}
\right).
\ee
Here (1,2,3,4,5) and (1,2,3) denote the cyclic permutations that generate the alternating group. We can now calculate the invariant tensor using these matrices, for example, we can use the Mathematica function ``NullSpace[ ]''.  Notice that the generators from ``IrreducibleRepresentationsDixon'' are not necessarily orthogonal. The bi-linear form 
\be
\Omega=\left(
\begin{array}{ccccc}
 1 & \frac{1}{3} & -\frac{1}{3} & \frac{1}{3} & -\frac{1}{3} \\
 \frac{1}{3} & 1 & -\frac{1}{3} & -\frac{1}{3} & -\frac{1}{3} \\
 -\frac{1}{3} & -\frac{1}{3} & 1 & \frac{1}{3} & -\frac{1}{3} \\
 \frac{1}{3} & -\frac{1}{3} & \frac{1}{3} & 1 & -\frac{1}{3} \\
 -\frac{1}{3} & -\frac{1}{3} & -\frac{1}{3} & -\frac{1}{3} & 1 \\
\end{array}
\right)
\ee
that is preserved by $g_1$ and $g_2$ is not proportional to the identity matrix. To fix this, we perform the Cholesky Decomposition of $\Omega=L^T L$. Then the new representation of the generators in terms of the orthogonal matrices is 
\be
g_1=(L^{T})^{-1} \tilde{g}_1 L,\quad  g_2=(L^{T})^{-1} \tilde{g}_2 L.
\ee
In this basis, we found the two degree three invariant polynomials to be 
\bea\label{poly5}
K_1&=&-\frac{3}{10} x \left(y \left(-5 t-\sqrt{15} w+2 \sqrt{15} z\right)+(2 w+z) \left(\sqrt{15} t-w+2 z\right)\right)+\nonumber\\&& \frac{\sqrt{2}}{200}\left(50 t^3+30 t w^2+12 \sqrt{15} w^3-12 w z \left(\sqrt{15} w-15 t\right)-21 z^2 \left(5 t+\sqrt{15} w\right)-\right.\nonumber\\
&&\left.15 y^2 \left(-5 t-\sqrt{15} w+2 \sqrt{15} z\right)-30 y (2 w+z) \left(\sqrt{15} t-w+2 z\right)+6 \sqrt{15} z^3\right),
\eea
and
\bea\label{poly52}
K_2&=&\frac{1}{10} x \left(-15 t^2+w^2+16 w z+5 y^2-11 z^2\right)+\frac{2}{3} x^3+\nonumber\\&&\frac{\sqrt{2}}{600}\left(250 y^3-15 y \left(-15 t^2+4 z \left(7 w-3 \sqrt{15} t\right)+6 \sqrt{15} t w+43 w^2+22 z^2\right)+\right.\nonumber\\&&
\left.9 (2 w+z) \left(-5 \sqrt{15} t^2-4 z \left(\sqrt{15} w-5 t\right)-10 t w+\sqrt{15} w^2+4 \sqrt{15} z^2\right)\right).
\eea
\subsubsection{The group S5}
The permutation group of five elements has two five dimensional irreps denoted by the Young diagram 
\ydiagram{3,2}
$ $ and 
\ydiagram{2,2,1}. For simplicity, we will denote them as $5_a$ and $5_b$. The group A5 can be embedded in S5 through the standard way. That is, S5 contains all the permutations of five elements, while A5 contains even permutation that permutes the five elements. Under this embedding, both $5_a$ and $5_d$ branch into the 5 dimensional irrep of A5. 
\bea
5_a \rightarrow 5,\nonumber\\
5_b \rightarrow 5.
\eea
One can similarly work out the invariant polynomials by using the generators from “IrreducibleRepresentationsDixon”. It turns out that the invariant polynomial of $5_a$ is related to $K_1$ by an O(5) rotation, while the invariant polynomial of $5_b$ is related to $K_2$ by an O(5) rotation.
\subsubsection{The group A6 and S6}
The group S6 contains all permutations of six elements. There are two 5 dimensional irreps that preserve a degree-three polynomial. One of them is the standard irreps of permutation groups ``$5_{std}$''. We will denote the other irrep as ``$5'$''. 
As is clear from Table \ref{group5inv}, the two irreps in fact have the same Molien series. This is only possible if the two irreps have exactly the same invariant polynomials. Physically, this means a potential $V(\phi)$ with $\phi$ transforms in the $5_{std}$ irrep can be also interpreted as a potential with $\phi$ transforms in the $5'$ irrep. 

The symmetric group S5 can be embedded in S6 in the standard way. While S6 contains all the permutation that permutes six elements, S5 contains all the permutation that permutes five of the six elements. Under this embedding 
\bea
5_{std} &\rightarrow& 4_{std}+1,\nonumber\\
 5' &\rightarrow& 5_{a}.
\eea
An invariant polynomial of the parent group S6 should also be an invariant polynomial of the subgroup S5. We, therefore, know the degree three invariant polynomial of $5'$ is simply $K_1$. Since $5_{std}$ and $5'$ have the same Molien series, the invariant polynomial of $5_{std}$ should also be $K_1$.

From Table \ref{group5inv}, we know the Molien series of the group A6 and S6 are the same up to $z^4$ terms. Physically, this means we can not break the symmetry from S6 to A5 until we introduce $\Phi^5$ terms in the (super-)potential. Since we consider only cubic terms of the superpotential, we should consider only the group S6.
\subsection{The beta functions and fixed points}\label{N5points}
We can now consider the beta function of the $\cal N$=1 Lagrangian with superpotential 
\be
\mathcal{W}= g_1 K_1+ g_2 K_2,
\ee
with $K_1$ and $K_2$ defined in \eqref{poly5} and \eqref{poly52}. The beta function is
\bea
\beta_{1}(g_1,g_2)&=&-\frac{\epsilon}{2}g_1+\frac{1}{16\pi^2} 3 g_1^3,\nonumber\\
\beta_{2}(g_1,g_2)&=&-\frac{\epsilon}{2}g_2+\frac{1}{16\pi^2}\frac{3g_2^3}{5}.
\eea
and the corresponding anomalous dimension is 
\be
\gamma_{\Phi}/\epsilon=\frac{1}{2}(g_1^2+g_2^2)
\ee
We find the following four fixed points:

A fixed point at ($\frac{g_1}{4\pi\sqrt{\epsilon}}=0,\frac{g_2}{4\pi\sqrt{\epsilon}}=\pm\sqrt{\frac{3}{10}}$) preserves S5 symmetry. It has $\gamma_{\Phi}/\epsilon=\frac{3}{20}$ and $A/\epsilon=-21 \pi ^2$. 

A fixed point at ($\frac{g_1}{4\pi\sqrt{\epsilon}}=\pm \frac{1}{\sqrt{6}},\frac{g_2}{4\pi\sqrt{\epsilon}}=0$) preserves S6 symmetry. It has $\gamma_{\Phi}/\epsilon=\frac{1}{12}$ and $A/\epsilon^2=-\frac{1}{3} \left(35 \pi ^2\right)$.

A fixed point at ($\frac{g_1}{4\pi\sqrt{\epsilon}}=\pm \frac{1}{\sqrt{6}},\frac{g_2}{4\pi\sqrt{\epsilon}}=\pm\sqrt{\frac{3}{10}}$) instead of the A5 symmetry, preserves O(3)/Z2 symmetry. The superpotential takes the form 
\be
\mathcal{W}=\sqrt{\frac{7}{15}}\left( \sqrt{\frac{5}{14}} K_1+\frac{3}{\sqrt{14}} K_2  \right),
\ee
The polynomial
\be
\sqrt{\frac{5}{14}} K_1+\frac{3}{\sqrt{14}} K_2 
\ee
is related to the invariant polynomial of O(3)/Z2 by an O(5) rotation. It has $\gamma_{\Phi}/\epsilon=\frac{7}{30}$ and $A/\epsilon^2=-\frac{1}{3} \left(98 \pi ^2\right)$. This fixed point is the most stable one. These A5 invariant RG flows have an IR fixed point with emergent O(3)/Z2 symmetry.

\section{Generalization to higher $N_{\Phi}$}\label{generalN}

\subsection{SN$\times$ SN bi-standard fixed points}
We discuss here the irreducible fixed points first. The decoupled S3 Potts model studied in Section \ref{S5potts} can also be understoodd as an SCFT with the four scalar superfields $\Phi_i$ transforming in the bi-standard representation of  S3$\times$ S3. This SCFT can be easily generalized to a higher number of scalar superfields, if we take the scalars to transform in the $(r_1,\ldots,r_n) $ representation of $G_1\times \ldots G_n$. 
The constant $T_3$ defined in \eqref{T3} is simply the product of the $T_3$'s of the individual $G$'s,
\be
T_3=T_3^{(G_1)}\times \ldots T_3^{(G_n)}.
\ee
Interestingly, if we take $G_1=G_1=\ldots G_n=G$, and if $n$ is an even number, 
\be
T_3\geq0,
\ee
The one-loop $\cal{N}$=1 SUSY fixed point is guaranteed to exist. Take the group to be SN$\times$ SN, we have then $T_3=(\frac{\rm N-3}{\rm N-2})^2$, so that
\be
\gamma_{\Phi}=\frac{\rm (N-2)^2  }{\rm 2 N (7 N-36)+96}\epsilon\,.
\ee
Also the A-function is 
\be
A/\epsilon^2=-28 \pi ^2\frac{\rm (N-2)^2 (N-1)^2 }{\rm 2 N (7 N-36)+96}\,.
\ee

\subsection{O(N)/Z2 fixed points}\label{ONTfixedpoints}

We now try to generalize the O(3)/Z2 fixed studied in Section \ref{O3Z2} to a family of fixed points with O(N)/Z2 symmetry. 
The scalar superfields $\Phi_{ij}$ carry indices in the T irrep of O(N), so that these fixed points have $N_{\phi}=\frac{\rm N(N+1)}{2}-1$ interacting scalar superfields.
The product of two vector irreps of O(N) can be decomposed into the S, A and T irreps:
\be
\imineq{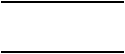}{6}=\quad \frac{1}{\rm N}\imineq{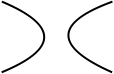}{8} \nn
+\quad \imineq{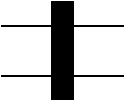}{11}\nn
+\imineq{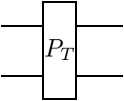}{11}\,.
\ee
The projector to the T irrep is
\be
\imineq{PT.png}{11}=\imineq{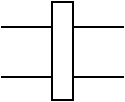}{11}-\frac{1}{\rm N}\imineq{singlet.png}{9}\,.
\ee
Here we use the birdtrack notation \cite{Cvitanovic:2008zz}. Solid lines mean the Kronecker $\delta_{ij}$, the unfilled box means symmetrization, and the filled box means anti-symmetrization. The invariant tensor for the T irrep of O(N) is simply, 
\be
\imineq{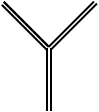}{11}=\imineq{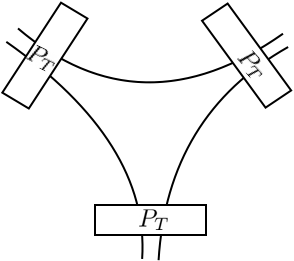}{17}\,.
\ee
We rescale the invariant tensor to satisfy the normalization \eqref{nomal}, which can now be represented as
\be
\imineq{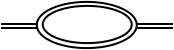}{6}\quad =\quad \imineq{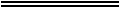}{1}\quad\,.
\ee
After contracting the Kronecker delta's, we get
\be
\imineq{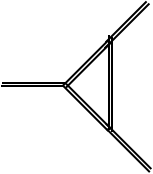}{17}=T_3 \quad \imineq{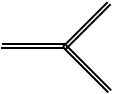}{9}\,,
\ee
with 
\be
T_3=\frac{\rm N^2+4 N-24}{\rm 2 (N^2+2 N-8)}\,.
\ee
Notice $T_3\geq-3/14$ for $N\geq3$, which satisfies the condition \eqref{condtion}. So that the fixed points exist at one loop. 
The anomalous dimension is then
\be
\gamma_{\Phi}=\frac{\rm N^2+2 N-8}{\rm 2 \left(5 N^2+14 N-72\right)}\epsilon\,.
\ee
The A-function is 
\be
A/\epsilon^2=-\frac{\rm 7 \pi ^2 (N-1) N \left(N^2+2 N-8\right)}{\rm 5 N^2+14 N-72}\,.
\ee
As already mentioned in Section \ref{O3Z2}, since the scalar superfields $\Phi_{ij}$ carry an O(N) T index, the central element of O(N)
\be
\textrm{Z}=\textrm{Diag}\{-1,\ldots,-1\}\,,
\ee
acts trivially on $\Phi_{ij}$. The flavor symmetry of the SCFTs is then
\be
\textrm{O(N)/Z2}\,.
\ee
\subsection{Generalization of A5}
The group A5 is very interesting.
As far as we know, this is the first example of a group that has an irrep that satisfies the trace condition (preserves a single degree two polynomial) and preserves more than one degree three polynomial. To generalize this to the higher-component case we use the birdtrack technique \cite{Cvitanovic:2008zz}. To be more specific,  we search for subgroups of SN groups that have an irrep with the Molien series 
\be
1+z^2+2z^3+2 z^4+O\left(z^5\right)\,.
\ee
Again we decompose two vector irreps of O(n) as S, A and T. If a subgroup of O(n) preserves a rank three symmetric traceless tensor, the T irrep of O(n) gets further decomposed into two irreps:
\bea
\imineq{twoline.png}{7}&=&\quad \frac{1}{n}\imineq{singlet.png}{9} \nn\\&&
+\quad \imineq{A.png}{11}\nn\\&&
+\quad \imineq{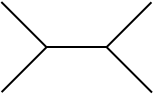}{9}\nn\\&&
+\quad \left(\imineq{symirep.png}{11}-\frac{1}{n}\imineq{singlet.png}{9}-\imineq{vecdecompose.png}{9}\right).
\eea
The invariant tensor of the standard irrep of SN satisfies a special condition. That is, the A irrep of O(n) does not decompose. In term of birdtracks, it implies the following identity:
\be
\imineq{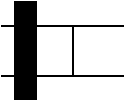}{11}=-\frac{1}{n-1} \imineq{A.png}{11}\,.
\ee
Here $$n={\rm N-1}\,,$$ is the dimension of the standard irrep of SN. The coefficient is fixed by contracting the top two legs. 
From the above relation we can show that the invariant tensor of SN satisfies
\be
\imineq{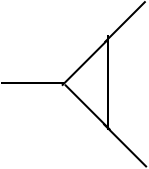}{13}=T_3\quad \imineq{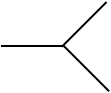}{9}\,,
\ee
with
\be
T_3=\frac{n-2}{n-1}\,.
\ee
We now want to generalize the group A5 to a group with higher $n$. According to the Molien series, we first introduce another invariant tensor 
\be
\imineq{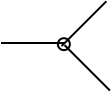}{9}\,.
\ee
The Molien series also tells us that the irrep has only two degree four invariant polynomials. This means the two invariant tensors satisfy
\be\label{4sym1}
\imineq{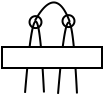}{11}=A\quad  \imineq{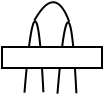}{11}+ A'\quad  \imineq{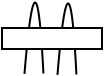}{8.5}\,,
\ee
and
\be\label{4sym2}
\imineq{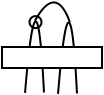}{11}=B\quad  \imineq{4syma.png}{11}+ B'\quad  \imineq{4symb.png}{8.5}\,.
\ee
Contracting two legs with $\delta_{ij}$, we get 
\be
A'=\frac{2 (1-A)}{n+2}\,, \quad 
B'=\frac{2 (-B)}{n+2}\,.
\ee
Where we have used the normalization conditions,
\bea\label{normalizations}
\imineq{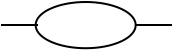}{5}&=& \imineq{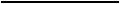}{0.4}\,,\nn\\
\imineq{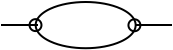}{5}&=& \imineq{line.png}{0.4}\,,\nn\\
\imineq{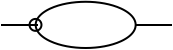}{5} &=& 0\,.
\eea

Contracting two legs of \eqref{4sym1} with $\imineq{v1.png}{9}$, we get 
\be\label{vcontraction1}
\imineq{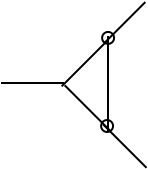}{13}=M\quad \imineq{v1.png}{9} .
\ee
with 
\be\label{aMrelation}
M=\frac{A n-2 A+4}{2 n+4}+A T_3\,.
\ee{}
From this we can prove 
\be\label{vcontraction2}
\imineq{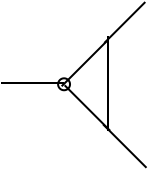}{13}=M\quad \imineq{v2.png}{9}.
\ee
The proof is given in Appendix \ref{proof}.
Contracting two legs of \eqref{4sym1} with $\imineq{v2.png}{9}$, we get
\be\label{vcontraction3}
\imineq{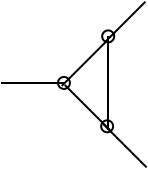}{13}= P\quad \imineq{v2.png}{9}
\ee
with 
\be
P=\frac{\left(A^2-1\right) (n-2)}{2 (n+2)}+A^2 T_3\,.
\ee
Contracting two legs of \eqref{4sym2} with $\imineq{v1.png}{9}$, we obtain
\be
\frac{2 B}{3 (n+2)}+\frac{1}{6} (4 M+1) \imineq{v1.png}{11}=(1/3) (1 + 2T_3)B\imineq{v2.png}{11}\,. 
\ee{}
Since the two tensor are independent, we get
\be
M=-\frac{1}{4} \quad \text{and}\quad B=0\,.
\ee{}
According to \eqref{aMrelation}, this leads to
\be
A=-\frac{(n-1) (n+10)}{6 (n-2) (n+1)}\,.
\ee
Using \eqref{4sym1} and \eqref{4sym2},  after a long calculation, we get 
\be
\imineq{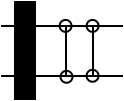}{11}=\frac{5 n^2-71 n+86}{12 (n-2) (n+1)} \imineq{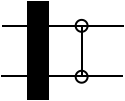}{11}+ \frac{66-15 n}{-8 n^2+8 n+16}\imineq{A.png}{11}\,.
\ee
From this we know how the anti-symmetric subspace can be decomposed using the projectors:
\be
P_1=\frac{(n+1) (5 n-22)}{(n+10) (5 n-13)} \imineq{A.png}{11}+\frac{12 \left(-n^2+n+2\right)}{(n+10) (5 n-13)}\imineq{Q.png}{11}\,,
\ee
and 
\be
P_2=\frac{54 (n-2)}{(n+10) (5 n-13)}\imineq{A.png}{11}+\frac{12 (n-2) (n+1)}{(n+10) (5 n-13)}\imineq{Q.png}{11}\,.
\ee
They satisfy
\be
P_1 P_1 =P_1,\quad \text{and}\quad P_2 P_2 =P_2\,.
\ee
The dimension of the corresponding irrep is
\be
d_1=\frac{5 n^2-71 n+86}{12 (n-2) (n+1)},\quad d_2=\frac{66-15 n}{-8 n^2+8 n+16}\,.
\ee
We can calculate the diagram 
\be
\imineq{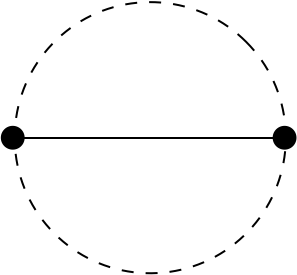}{12}=\imineq{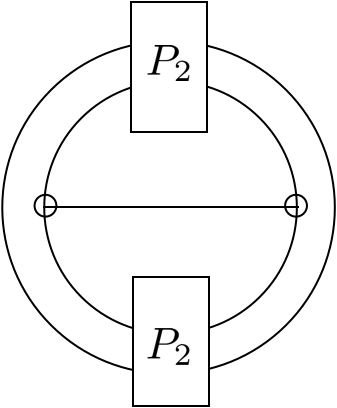}{15}=-\frac{n (7 n-11) (55 n-98) ((n-43) n+118)}{16 (13-5 n)^2 (n+10)^2}\,.
\ee
Here we use $\imineq{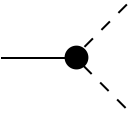}{8}$ to denote the Clebsch–Gordan coefficients $(T_i)_{ab}$ of $P_2\times P_2\rightarrow n$.
This diagram is a sum of squares, 
\be
\sum_{i,a,b}(T_i)_{ab}(T_i)_{ab}\,,
\ee and is therefore non-negative. This puts a constraint on the values that $n$ can take. Notice that $n$ is a positive integer, so that 
\be
3 \leq n\leq 40\,.
\ee
The dimensions of the irreps $d_1$ and $d_2$ also need to be positive integers, there are only two solutions of these Diophantine conditions, they are 
\bea
n=5, &&d_1=4\,, \quad\quad d_2=6\,,\nonumber\\
n=26, && d_1=249\,, \quad d_2=76\,.
\eea
The $n=5$ solution corresponds to the group A5 we have studied, while the $n=26$ solution might correspond to a subgroup of S27.
  
Let us assume that such a group exists and calculate the beta function and anomalous dimensions. 
Take the superpotential to be 
\be
\mathcal{W}=g_1 \imineq{v1.png}{11}+g_2 \imineq{v2.png}{11}\,,
\ee
using the birdtrack relations \eqref{vcontraction1}, \eqref{vcontraction2} and \eqref{vcontraction3}, we get
\bea
\beta_{1}(g_1,g_2)&=&-\frac{\epsilon}{2}g_1+\frac{1}{16\pi^2} \frac{171}{50} g_1^3\,,\nonumber\\
\beta_{2}(g_1,g_2)&=&-\frac{\epsilon}{2}g_2+\frac{1}{16\pi^2}\frac{19g_2^3}{24}\,,
\eea
and the corresponding anomalous dimension is 
\be
\gamma_{\Phi}/\epsilon=\frac{1}{2}(g_1^2+g_2^2)\,.
\ee
We find the following fixed points:

A fixed point at ($\frac{g_1}{4\pi\sqrt{\epsilon}}=0,\frac{g_2}{4\pi\sqrt{\epsilon}}=\pm2 \sqrt{\frac{3}{19}}$) has $\gamma_{\Phi}/\epsilon=\frac{6}{19}$ and $A/\epsilon^2=-\frac{1}{19} \left(4368 \pi ^2\right)$. 

A fixed point at ($\frac{g_1}{4\pi\sqrt{\epsilon}}=\pm \frac{5}{3 \sqrt{19}},\frac{g_2}{4\pi\sqrt{\epsilon}}=0$) preserves S27 symmetry. It has $\gamma_{\Phi}/\epsilon=\frac{25}{342}$ and $A/\epsilon^2=-\frac{1}{171} \left(9100 \pi ^2\right)$.

A fixed point at ($\frac{g_1}{4\pi\sqrt{\epsilon}}=\pm \frac{5}{3 \sqrt{19}},\frac{g_2}{4\pi\sqrt{\epsilon}}=\pm2 \sqrt{\frac{3}{19}}$) has $\gamma_{\Phi}/\epsilon=\frac{7}{18}$ and $A/\epsilon^2=-\frac{1}{9} \left(2548 \pi ^2\right)$.

Notice the birdtrack rules \eqref{4sym1} and \eqref{4sym2} are very similar to the rules that lead to the classification of the F4-family of Lie groups, see \cite{Cvitanovic:2008zz} equation (19.16). Other birdtrack conditions lead to the classification of E6, E7 and E8 family of Lie Groups. Deligne conjectured that there exist categories interpolating these exceptional groups \cite{deligne1996serie, deligne2002exceptional}. See \cite{Binder:2019zqc} for how to use the Deligne category to make sense of O(N) invariant quantum field theories at non-integer N. It is tempting to conjecture that \eqref{4sym1} and \eqref{4sym2} also lead to a new Deligne category. We leave this for future work.

\subsection{The $\mathcal{N}$=1 Potts models}
The $\mathcal{N}$=1 Potts models were already studied in \cite{Rong:2019qer}. We recall here the procedure to construct the invariant tensor $d_{ijk}$ according to \cite{Zia:1975ha}.  In \cite{Rong:2019qer}, the $d_{ijk}$ were constructed explicitly and used to calculated $\gamma_{\Phi}$ to two loops in $\epsilon$. The result was shown to be consistent with the result from the non-perturbative bootstrap. 
Take the scalar superfield $\Phi_i$ to transform in the ${\rm N}-1$ dimensional standard representation of SN. First, we need to construct $e^{\alpha}_i$ with $\alpha=1\ldots {\rm N}$ and $i=1\ldots {\rm N}-1$. The $e^{\alpha}$ is a vector in the ${\rm N}-1$ dimensional Euclidean space $\mathbb{R}^{{\rm N}-1}$. The index $\alpha$ labels the ${\rm N}$ vertices of a hypertetrahedron. The $e^{\alpha}_i$ can be constructed through a recursion relation,
\bea
&e_i^{\alpha}&= \sin(\theta_{{\rm N}-1})\tilde{e}_i^{\alpha} \quad\quad \textrm{for} \quad\alpha=1\ldots { \rm N-1};\quad i=1\ldots {\rm N}-2\,,\nn\\
&e_N^{\alpha}&= \cos(\theta_{{\rm N}-1}) \quad\quad\quad \textrm{for}\quad  \alpha=1\ldots {\rm N}-1\,,\nn\\
&e_i^{N}&=\delta_{{\rm N}-1,i}\,,
\eea
with $\cos(\theta_{{\rm N}-1})=-1/{({\rm N}-1)}$.
Here $\tilde{e}^{\alpha}_{i} $ is the set of N-1 vectors in  $\mathbb{R}^{{\rm N}-2}$. The recursion starts with $e^1=\{1\}$ and $e^2=\{-1\}$.
The invariant tensor $d_{ijk}$ is defined as,
\be\label{Pottsinv}
d_{ijk}=\sqrt{\frac{\rm (N-1)^3}{\rm (N-2) N^2}}\sum_{\alpha=1}^{\rm N}e_i^{\alpha}e_j^{\alpha}e_k^{\alpha}\,.
\ee
The overall constant is chosen so that the invariant tensor satisfies the normalization \eqref{nomal}. From the above definition of $d_{ijk}$, we get $$T_3=\frac{\rm N-3}{\rm N-2}\,,$$
\be
\gamma_{\Phi}=\frac{\rm \rm N-2}{\rm 14 N-36}\epsilon\,,
\ee
and 
\be
A/\epsilon^2=-\frac{\rm 28 \pi ^2 (N-2) (N-1)}{\rm 14 N-36}\,.
\ee

\subsection{Generalization of reducible fixed points}
Take the superpotential to be
\be
W=\lambda_1 H \sum_{i=1}^{\rm N-1}\Phi_i\Phi_i+\lambda_2 d_{ijk}\Phi_i\Phi_j\Phi_k+\lambda_3H^3\,.
\ee

The $\beta$ functions for the three coupling turn out to be 
\bea\label{threecouplingbeta}
\beta_{\lambda_1}&=&-\frac{\epsilon}{2}\lambda_1+ \frac{1}{(4 \pi )^2}\frac{1}{18} \left(7 \lambda _1^3+12 \lambda _3 \lambda _1^2+54 \lambda _2^2 \lambda _1+9 \lambda _3^2 \lambda _1+\lambda _1^3 {\rm N}\right)\,,\nonumber\\
\beta_{\lambda_2}&=&-\frac{\epsilon}{2}\lambda_2+\frac{1}{(4 \pi )^2}\frac{1}{2} \left(3 \lambda _2^3+2 \lambda _1^2 \lambda _2+4 \lambda _2^3 T_3\right)\,,\nonumber\\
\beta_{\lambda_3}&=&-\frac{\epsilon}{2}\lambda_3+ \frac{1}{(4 \pi )^2}\frac{1}{54} \left(-4 \lambda _1^3-9 \lambda _3 \lambda _1^2+189 \lambda _3^3+4 \lambda _1^3 N+9 \lambda _3 \lambda _1^2 {\rm N}\right)\,.
\eea
The anomalous dimension is given by 
\bea
\gamma_{H}&=&\frac{1}{(4\pi)^2}\frac{1}{18} \left(9 \lambda _3^2+\lambda _1^2 ({\rm N}-1)\right)\,,\nonumber\\
\gamma_{\Phi}&=&\frac{1}{(4\pi)^2}\left(\frac{\lambda _1^2}{9}+\frac{\lambda _2^2}{2}\right)\,.
\eea
The constant $T_3$ is defined in \eqref{T3}. Take $d_{ijk}$ to be the invariant tensor of the standard irrep of SN \eqref{Pottsinv}.  
At 
\bea
\frac{\lambda_1}{4\pi\sqrt{\epsilon}}&=&-\frac{3}{\sqrt{\rm N (7 N-11)}}\,,\\
\frac{\lambda_2}{4\pi \sqrt{\epsilon}}&=&-\frac{\rm \sqrt{(N-2) (N+1) (7 N-18)}}{\rm \sqrt{N (7 N-18) (7 N-11)}}\,,\\
\frac{\lambda_3}{4\pi \sqrt{\epsilon}}&=&\frac{\rm N-1}{\sqrt{\rm N (7 N-11)}}\,,
\eea
the scalar superfield $H$ and $\Phi_i$ together form the $\rm N$ dimensional standard representation of $S_{\rm N+1}$. This corresponds to the $S_{\rm N+1}$ Potts model studied in the previous subsection.

\subsubsection{The O(N-1) fixed points with with $\gamma_{H}\neq\gamma_{\Phi}$}
When $\lambda_2=0$, for any value of N, there exists only one perturbatively unitary O(N-1) invariant fixed point with real coupling constant. This has been studied in \cite{Benini:2018bhk}. The scalar superfields $\Phi_i$ transform in the vector representation of O(N-1) while $H$ is invariant under an O(N-1) transformation. 
The value of the coupling at the fixed can be worked out analytically. It is however rather complicated and not very illuminating. Let us record here the large N result,
\bea
\frac{\lambda_1}{4\pi\sqrt{\epsilon}}&=&\frac{1}{\rm \sqrt{\rm N}}\left( -3+\frac{9}{2 \rm N}-\frac{81}{8 {\rm N}^2}+\mathcal{O}(\frac{1}{\rm N^3})\right)\,,\nonumber\\
\frac{\lambda_3}{4\pi\sqrt{\epsilon}}&=&\frac{1}{\rm \sqrt{\rm {\rm N}}}\left(2-\frac{27}{\rm {\rm N}}+\frac{3915}{4 {\rm N}^2} +\mathcal{O}(\rm \frac{1}{\rm N^3})\right)\,.
\eea
The anomalous dimension is given by
\bea
\gamma_{H}/\epsilon&=&\frac{1}{2}-\frac{48}{{\rm N}^2}+\frac{36999}{16 {\rm N}^3}+\mathcal{O}(\frac{1}{N^3})\,,\nonumber\\
\gamma_{\Phi}/\epsilon&=&\frac{1}{\rm N}-\frac{3}{\rm N^2}+\frac{9}{\rm N^3}+\mathcal{O}(\frac{1}{\rm N^3})\,.
\eea
The A-function is 
\be
A/\epsilon^2=-42 \pi ^2+\frac{112 \pi ^2}{\rm N}+\frac{1008 \pi ^2}{\rm N^2}-\frac{257985 \pi ^2}{4 \rm N^3}+\mathcal{O}(\frac{1}{\rm N^3})\,.
\ee
In three dimensions, the large N limit of the fixed points is dual to  $\mathcal{N}$=1 higher-spin
theory on AdS$_4$ through AdS/CFT correspondence \cite{Leigh:2003gk,Sezgin:2003pt}. 

\subsubsection{SN fixed point with $\gamma_{H}\neq\gamma_{\Phi}$ }
The fixed point is located at 
\bea
\frac{\lambda_1}{4\pi\sqrt{\epsilon}}&=&-\frac{18 \sqrt{\rm N-6}}{\sqrt{\rm N (N (7 N-18)-864)}}\,,\\
\frac{\lambda_2}{4\pi \sqrt{\epsilon}}&=&\frac{\sqrt{\rm (N-2) (7 N-18) \left(N^2-216\right)}}{\sqrt{\rm N (7 N-18) (N (7 N-18)-864)}}\,,\\
\frac{\lambda_3}{4\pi \sqrt{\epsilon}}&=&\frac{\rm (N-6)^{3/2}}{\rm \sqrt{N (N (7 N-18)-864)}}\,.
\eea
(There are three equivalent fixed points related to the above one by field re-definition.) The scalar superfields $\Phi_i$ transform in the N-1 dimensional standard representation representation of SN while $H$ is invariant under SN transformation. The anomalous dimension is given by
\be
\gamma_{H}=\frac{\rm (N-6) (N+24)}{2 (N (7 N-18)-864)}\epsilon\,,\quad \gamma_{\Phi}=\frac{\rm (N-2) N-144}{\rm 2 (N (7 N-18)-864)}\epsilon\,.
\ee
The A-function is 
\be
A/\epsilon^2=-\frac{\rm 14 \pi ^2 N ((N-2) N-124)}{\rm N (7 N-18)-864}\,.
\ee
Notice there exists a non-conformal window given by
\be
\rm 6<N<14.7\,,
\ee
within which the fixed point is non-unitary. The upper edge of the window is $6\sqrt{6} \approx 14.7$.
At the edges of the window 
\bea
{\rm N}_{c,\text{lower}}&=&6: \quad\quad \lambda_1=0,\quad \lambda_2=\frac{1}{\sqrt{6}},\quad \lambda_3=0\,,\nonumber\\
{\rm N}_{c,\text{upper}}&=&6\sqrt{6}:\quad \lambda_1=\frac{1}{\sqrt{2}}\,,\quad \lambda_2=0,\quad \lambda_3=\frac{1}{6} \left(\sqrt{2}-2 \sqrt{3}\right)\,. 
\eea
Clearly, at the lower edge of the non-conformal window, the fixed point collides with the decoupled SN Potts models and one free scalar theory. At the upper edge of the non-conformal window the fixed point collides with the O(N-1) invariant fixed point.
\begin{figure}[ht]
\includegraphics[width=10cm]{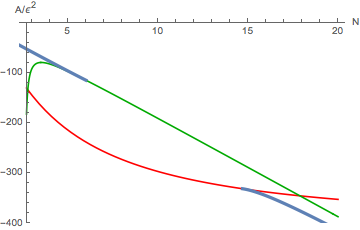}
\caption{$A$ function for the O(N-1) fixed point (red), the SN invariant Potts model+free superfield (green), and the new SN invariant fixed point (blue). The new SN invariant fixed point goes to the complex plane after collision with the other fixed points. The RG flow is therefore not conformal in the IR, leading to a non-conformal window.  }
\label{island}
\end{figure}

The beta functions \eqref{threecouplingbeta} have a few other fixed points. They are,  however, decoupled combinations of either the super-Ising model, SN Potts model, or the free theory. We will not list them here.

\section{Discussion}

A natural generalization of the $\mathcal{N}$=1 analysis of this paper is to study $\mathcal{N}$=2 SCFTs. Consider the  following $\mathcal{N}$=2 superpotential
\be\label{N2Superpotential}
\mathcal{W}=\frac{1}{6}h^{ijk}\Phi_i\Phi_j\Phi_k\,.
\ee
The $\beta$ function is given by
\be
\beta_{abc}=-\frac{\epsilon}{2}h^{ijk}+\gamma^{i}{}_{m}h^{mjk}+\gamma^{j}{}_{m}h^{mik}+\gamma^{k}{}_{m}h^{ijm}\,,
\ee
and the anomalous dimension matrix reads 
\be
\gamma^{i}{}_{m}=\frac{1}{2} h^{iab}h_{mab}\,.
\ee
The scalar superfield $\Phi$ is complex so that the coupling $h^{ijk}$ transform in the fully symmetric representation of U(N). By studying subgroups of U(N) and analyzing their degree three invariant polynomials we can study the corresponding $\mathcal{N}$=2 superconformal fixed points. 
Among the 109 finite subgroups of U(5) with 5 dimensional faithful irreducible representations, we found a single group whose Molien Series takes the form
\be
1+2 z^3+O\left(z^4\right)\,.
\ee
The group has a SmallGroup id [180,19] and structure description GL(2,4).
This is the general linear group of degree two over the field of four elements. 
Notice that the corresponding irrep has no degree two invariant polynomials, it is in fact a complex irrep.  This irrep has two degree three polynomials. 
Using these two polynomials to construct an $\mathcal{N}$=2 superpotential as in \eqref{N2Superpotential}, and analyzing the beta function, we get a full  conformal manifold.
The group GL(2,4) is in fact the direct product of A5 and cyclic group of order three Z3. The invariant polynomials of GL(2,4) are simply $K_1$ and $K_2$ as defined in \eqref{poly5} and \eqref{poly52}. The five variables $\{x,y,z,w,t\}$ are now complex. The action of Z3 is 
\be
\{x,y,z,w,t\} \rightarrow e^{\frac{2\pi \im}{3}}\{x,y,z,w,t\}\,.
\ee
It might be interesting to study this conformal manifold in more detail. A similar study of a conformal manifold containing the XYZ model was performed in \cite{Baggio:2017mas}.

Notice the degree three invariant polynomials studied in this paper can be used to construct a Landau theory with cubic terms. The Landau criterion says that a second-order phase transition is possible only if the irrep of 
the order parameter 
has no degree three invariant polynomials. This however is known to fail in two dimensions. For example, the 3 state Potts model is known to go through a second order phase transition at the critical temperature. The corresponding Landau theory with a cubic term has an infra-red conformal fixed point. It will be interesting to understand whether the Landau theories analogous to the SCFTs studied in this paper lead to unitary conformal field theories in two dimensions. The problem perhaps can be studied using conformal bootstrap techniques. 

Due to the special form of the superpotential \eqref{N1superpotential}, we focus here on finite subgroups of O(4) and O(5) which preserve degree three invariant polynomials. 
It will also be interesting to study degree four polynomials, which can then be used to construct $\lambda \phi^4$ theories in $4-\epsilon$ dimensions. The early works in the 80's focused on $\lambda \phi^4$ theories where the scalars form a single irreducible representation of the symmetry group. 
In \cite{michel1981landau,toledano1985renormalization}, all the subgroups of O(4) and the corresponding $\lambda \phi^4$ theories were studied extensively.
The work of  \cite{hatch1985selection,kim1986classification,hatch1986renormalization,stokes1987continuous} studied the $\lambda \phi^4$ theories with N=6 and N=8 scalars forming irreducible representations of the 230 crystallographic space groups (see also \cite{stokes1988isotropy}). 
Recently, there was also revived interest on fixed points of reducible $\lambda \phi^4$ theories where the scalars form more than one irrep of the symmetry groups \cite{Osborn:2017ucf,Rychkov:2018vya,Hogervorst:2020gtc,Osborn:2020cnf}. Some of these perturbative fixed points were shown to survive in three dimensions using the numerical bootstrap \cite{Rong:2017cow, Stergiou:2018gjj,Kousvos:2018rhl,Stergiou:2019dcv,Kousvos:2019hgc,Henriksson:2020fqi,Henriksson:2021lwn}. 
A full classification of the irreducible $\lambda \phi^4$ theories with N scalars will need the classification of subgroups of O(N), this can be difficult when N is large. The method used in Section \ref{GAPsubgroups} based on the GAP system can be easily generalized to study $\lambda \phi^4$ theories. It will be interesting to see whether we can discover new fixed points in the $4-\epsilon$ expansion. 

As mentioned in the introduction, the numerical bootstrap method has proven to be very powerful in studying $2+1$ dimensional $\mathcal{N}$=1 SCFTs. In particular, the work of \cite{Rong:2019qer} studied three infinite families of Wess-Zumino models. It will be interesting to apply the numerical bootstrap to the new O(N)/Z2 family discovered in this work. 
As usual, due to the non-perturbative nature of the numerical bootstrap, we expect rigorous numerical results that can then be compared to other methods such as the $\epsilon$-expansion. We leave this interesting analysis for future work.

\begin{table}[h]
\resizebox{\textwidth}{!}{%
\begin{tabular}{|l|l|}
\hline [ 55, 1 ], C11 $\rtimes$ C5 & *[ 60, 5 ], A5 \\\hline *[ 80, 49 ],
  (C2 $\times$ C2 $\times$ C2 $\times$ C2) $\rtimes$ C5 & [ 110, 2 ], C2 $\times$ (C11 $\rtimes$ C5)\\ \hline
 *[ 120, 34 ], S5 & *[ 120, 35 ], C2 $\times$ A5\\\hline [ 125, 3 ], (C5 $\times$ C5) $\rtimes$ C5 & [ 125, 4 ],   C25 $\rtimes$ C5\\\hline [ 155, 1 ], C31 $\rtimes$ C5 & *[ 160, 234 ],   ((C2 $\times$ C2 $\times$ C2 $\times$ C2) $\rtimes$ C5) $\rtimes$ C2\\\hline 
*[ 160, 235 ],  C2 $\times$ ((C2 $\times$ C2 $\times$ C2 $\times$ C2) $\rtimes$ C5) &  [ 165, 1 ], C3 $\times$ (C11 $\rtimes$ C5)\\\hline
  [ 180, 19 ], GL(2,4)& [ 205, 1 ], C41 $\rtimes$ C5\\\hline
 [ 220, 2 ], C4 $\times$ (C11 $\rtimes$ C5) &[ 240, 91 ], A5 $\rtimes$ C4\\\hline 
 [ 240, 92 ], C4 $\times$ A5 &  *[ 240, 189 ], C2 $\times$ S5\\\hline
 [ 240, 199 ], C3 $\times$ ((C2 $\times$ C2 $\times$ C2 $\times$ C2) $\rtimes$ C5) & 
  [ 250, 8 ], ((C5 $\times$ C5) $\rtimes$ C5) $\rtimes$ C2\\\hline [ 250, 10 ], C2 $\times$ ((C5 $\times$ C5) $\rtimes$ 
C5) &  [ 250, 11 ], C2 $\times$ (C25 $\rtimes$ C5) \\\hline [ 275, 1 ], C11 $\rtimes$ C25 & [ 275, 3 ], 
  C5 $\times$ (C11 $\rtimes$ C5)\\\hline 
[ 300, 22 ], C5 $\times$ A5 & [ 305, 1 ], C61 $\rtimes$ C5\\\hline
  [ 310, 2 ], C2 $\times$ (C31 $\rtimes$ C5). &[ 320, 1583 ],   ((C2 $\times$ C2 $\times$ C2 $\times$ C2) $\rtimes$ C5) $\rtimes$ C4\\\hline
[ 320, 1584 ], C4 $\times$ ((C2 $\times$ C2 $\times$ C2 $\times$ C2) $\rtimes$ C5) & *[ 320, 1635 ], 
  ((C2 $\times$ C2 $\times$ C2 $\times$ C2) $\rtimes$ C5) $\rtimes$ C4\\\hline *[ 320, 1636 ],   C2 $\times$ (((C2 $\times$ C2 $\times$ C2 $\times$ C2) $\rtimes$ C5) $\rtimes$ C2)& [ 330, 4 ], C6 $\times$ (C11 $\rtimes$ C5)\\\hline
  [ 355, 1 ], C71 $\rtimes$ C5  & *[ 360, 118 ], A6\\\hline
 [ 360, 119 ], C3 $\times$ S5 &  [ 360, 122 ], C6 $\times$ A5 \\\hline
  [ 375, 2 ],((C5 $\times$ C5) $\rtimes$ C5) $\rtimes$ C3 & [ 375, 4 ], C3 $\times$ ((C5 $\times$ C5) $\rtimes$ C5)\\\hline   [ 375, 5 ], C3 $\times$ (C25 $\rtimes$ C5) & [ 385, 1 ],   C7 $\times$ (C11 $\rtimes$ C5) \\\hline
 [ 400, 52 ], (C2 $\times$ C2 $\times$ C2 $\times$ C2) $\rtimes$ C25 &  [ 400, 213 ],   C5 $\times$ ((C2 $\times$ C2 $\times$ C2 $\times$ C2) $\rtimes$ C5) \\\hline
 [ 405, 15 ], (C3 $\times$ C3 $\times$ C3 $\times$ C3) $\rtimes$ 
C5 &   [ 410, 2 ], C2 $\times$ (C41 $\rtimes$ C5) \\\hline 
[ 420, 13 ], C7 $\times$ A5 &
 [ 440, 2 ],  C8 $\times$ (C11 $\rtimes$ C5) \\\hline  [ 465, 2 ], C3 $\times$ (C31 $\rtimes$ C5) & 
[ 480, 217 ], A5 $\rtimes$ C8  \\\hline    [ 480, 220 ], C8 $\times$ A5 &
 [ 480, 943 ], C4 $\times$ S5  \\\hline [ 480, 1194 ] ,
  C3 $\times$ (((C2 $\times$ C2 $\times$ C2 $\times$ C2) $\rtimes$ C5) $\rtimes$ C2) & 
 [ 480, 1204 ],   C6 $\times$ ((C2 $\times$ C2 $\times$ C2 $\times$ C2) $\rtimes$ C5)\\\hline
[ 495, 1 ], C9 $\times$ (C11 $\rtimes$ C5) &  [ 500, 11 ], ((C5 $\times$ C5) $\rtimes$ C5) $\rtimes$ C4 \\\hline
[ 500, 13 ], C4 $\times$ ((C5 $\times$ C5) $\rtimes$ C5) &    
[ 500, 14 ], C4 $\times$ (C25 $\rtimes$ C5)  \\\hline
[ 500, 25 ], ((C5 $\times$ C5) $\rtimes$ C5) $\rtimes$ C4 &
   [ 500, 33 ], C2 $\times$ (((C5 $\times$ C5) $\rtimes$ C5) $\rtimes$ C2) \\\hline
[ 505, 1 ], C101 $\rtimes$ C5 &    [ 540, 31 ], C9 $\times$ A5\\\hline
 [ 550, 2 ]
C2 $\times$ (C11 $\rtimes$ C25) & 
[ 550, 10 ],   C10 $\times$ (C11 $\rtimes$ C5) \\\hline 
 [ 560, 173 ]  C7 $\times$ ((C2 $\times$ C2 $\times$ C2 $\times$ C2) $\rtimes$ C5)& 
  [ 600, 144 ], C5 $\times$ S5 \\\hline
 [ 600, 147 ], C10 $\times$ A5 &  [ 605, 1 ], C121 $\rtimes$ C5\\\hline
   [ 605, 3 ], C11 $\times$ (C11 $\rtimes$ C5) & [ 605, 5 ], (C11 $\times$ C11) $\rtimes$ C5 \\\hline
 [ 605, 6 ],   (C11 $\times$ C11) $\rtimes$ C5 &[ 610, 2 ], C2 $\times$ (C61 $\rtimes$ C5)\\\hline
 [ 615, 1 ],   C3 $\times$ (C41 $\rtimes$ C5) & [ 620, 2 ]C4 $\times$ (C31 $\rtimes$ C5)\\\hline
[ 625, 6 ], C125 $\rtimes$ C5. &  [ 625, 7 ], (C5 $\times$ C5 $\times$ C5) $\rtimes$ C5 \\\hline 
[ 625, 8 ], (C25 $\rtimes$ C5) $\rtimes$ C5 &  [ 625, 9 ], (C25 $\times$ C5) $\rtimes$ C5 \\\hline 
 [ 625, 10 ], (C25 $\times$ C5) $\rtimes$ C5 & [ 625, 14 ],   (C25 $\times$ C5) $\rtimes$ C5 \\\hline
 [ 640, 19097 ], ((C2 $\times$ C2 $\times$ C2 $\times$ C2) $\rtimes$ C5) $\rtimes$ C8 &
  [ 640, 19103 ], C8 $\times$ ((C2 $\times$ C2 $\times$ C2 $\times$ C2) $\rtimes$ C5) \\\hline
  [ 640, 21456 ], 
  ((C2 $\times$ C2 $\times$ C2 $\times$ C2) $\rtimes$ C5) $\rtimes$ C8 &[ 640, 21458 ], 
  C4 $\times$ (((C2 $\times$ C2 $\times$ C2 $\times$ C2) $\rtimes$ C5) $\rtimes$ C2) \\\hline
   *[ 640, 21536 ],   C2 $\times$ (((C2 $\times$ C2 $\times$ C2 $\times$ C2) $\rtimes$ C5) $\rtimes$ C4) & [ 655, 1 ], C131 $\rtimes$ C5 
 \\\hline [ 660, 4 ], C12 $\times$ (C11 $\rtimes$ C5) &
[ 660, 13 ], PSL(2,11) \\\hline [ 660, 14 ], 
  C11 $\times$ A5 & [ 710, 2 ], C2 $\times$ (C71 $\rtimes$ C5) \\\hline
 [ 715, 1 ], C13 $\times$ (C11 $\rtimes$ C5) & 
  [ 720, 399 ], C9 $\times$ ((C2 $\times$ C2 $\times$ C2 $\times$ C2) $\rtimes$ C5)\\\hline [ 720, 412 ], 
  C3 $\times$ (A5 $\rtimes$ C4) &[ 720, 419 ], C12 $\times$ A5 \\\hline
*[ 720, 763 ], S6 &
  *[ 720, 766 ], C2 $\times$ A6 \\\hline [ 720, 769 ], C6 $\times$ S5 & [750,6], ((C5 $\times$ C5) $\rtimes$ C5) $\rtimes$ C6 \\\hline [750,7], C2 $\times$ (((C5 $\times$ C5) $\times$ C5) $\rtimes$ C3) & [750,13], C3 $\times$ (((C5 $\times$ C5) $\rtimes$ C5) $\rtimes$ C2)\\\hline [750,24], C6 $\times$ ((C5 $\times$ C5) $\rtimes$ C5) & [750,25], C6 $\times$ (C25 $\rtimes$ C5)\\\hline [755,1], C151 $\rtimes$ C5& [770, 4], C14 $\times$ (C11 $\rtimes$ C5)\\\hline [775,1], C31 $\rtimes$ C25 & [780,13], C13 $\times$ A5\\\hline [800,384], C2 $\times$ ((C2 $\times$ C2 $\times$ C2 $\times$ C2) $\rtimes$ C25) & [800,1195], C5 $\times$ (((C2 $\times$ C2 $\times$ C2 $\times$ C2) $\rtimes$ C5) $\rtimes$ C2)\\\hline [800,1203], C10 $\times$ ((C2 $\times$ C2 $\times$ C2 $\times$ C2) $\rtimes$ C5) & \\\hline
\end{tabular}%
}
\caption{Finite subgroups of U(5) of order up to 800 with faithful 5-dimensional irreducible representations. Groups marked with ``*'' are also subgroups of O(5). The two integers [p,q] denote their id in Small Groups library, and p is the order of the finite group. }\label{groups}
\end{table}

\vskip 0.4 in
{\noindent\large  \bf Acknowledgments}
\vskip 0.in
J.~R. would like to thank Ning Su for discussions, and Andreas Stergiou for introducing to him the GAP system during the ``Bootstrap 2019'' conference held at Perimeter Institute.
The work of P.~L. and J.~R. is supported by the DFG through the Emmy
Noether research group ``The Conformal Bootstrap Program'' project number 400570283.

\appendix 
\section{Proof of \eqref{vcontraction3}}\label{proof}
Assume 
\be
\imineq{v12cub2.png}{13}=a\quad \imineq{v1.png}{9} +b \quad   \imineq{v2.png}{9} .
\ee
Contracting from the right with $\imineq{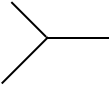}{9}$ and use \eqref{normalizations}, we get 
$a=0$. So that we have 
\be
\imineq{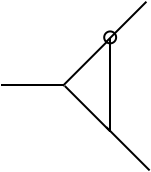}{13}=b\quad  \imineq{v2.png}{9}.
\ee
Contracting from the right with $\imineq{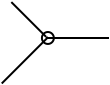}{9}$ and use \eqref{vcontraction1}, we get
\be
\imineq{v12cub2.png}{13}=M\quad \imineq{v2.png}{9}.
\ee
This completes the proof.
\bibliographystyle{JHEP}
\bibliography{reference.bib}

\end{document}